\documentclass[reprint]{revtex4-1}
\usepackage{graphicx}
\usepackage{dcolumn}
\usepackage{bm}

\begin{document}

\preprint{}

\title{Monte-Carlo simulation for the frequency comb spectrum of an atom laser}

\author{Dr.~Alexej~Schelle}
\affiliation{Freelancer Lecturer @ IU Internationale Hochschule, Juri-Gagarin-Ring 152, D-99084 Erfurt}

\date{\today}

\begin{abstract}
A theoretical particle-number-conserving quantum field theory based on the concept of imaginary time is presented and applied to the scenario of a coherent atomic laser field at ultra-cold temperatures. The proposed theoretical model describes the analytical derivation of the frequency comb spectrum for an atomic laser realized from modeling a coherent atomic beam of condensate and non-condensate quantum field components released from a trapped Bose-Einstein condensate at a given repetition phase and frequency. The condensate part of the atomic vapor is assumed to be subjected to thermal noise induced by the temperature of the surrounding thermal atomic cloud. This new quantum approach uses time periodicity and an orthogonal decomposition of the quantum field in a complex-valued quantum field representation to derive and model the quantum field's forward- and backward-propagating components as a standing wave field in the same unique time and temperature domain without quantitative singularities at finite temperatures. The complex-valued atom laser field, the resulting frequency comb, and the repetition frequency distribution with the varying shape of envelopes are numerically monitored within a Monte-Carlo sampling method, as a function of temperature and trap frequency of the external confinement.

\begin{description}
\item[Purpose]
Quanta 2023; 12: 171–179
\end{description}
\end{abstract}

\maketitle

\section{Introduction}

The theoretical derivation of the comb-like frequency distribution of a coherent laser obtained from a quantum field description usually relies on the mathematical concept of complex-valued Fourier transformation \cite{ref-0}. So far, standard theories for optical lasers have been applied mostly to model coherent optical lasers without thermal noise. Since the concept of Fourier transformation does not intrinsically describe the interplay between the coherent waves of the laser field and quantum noise effects induced by the thermal environment, it can therefore e. g. not be straightforwardly applied to the scenario of coherent atomic sources, where thermal and finite size effects on the coherence properties of the quantum wave field become important \cite{ref-1}. Using standard lasers, where amplification of the photons is realized by the stimulated emission of photons in the optical medium of a microwave cavity \cite{ref-2, ref-3}, coherent laser light in the absence of thermal noise and non-linear dispersion is built of Gaussian-shaped wave packets that can be outcoupled from the microwave cavity with a certain repetition frequency $\omega$. Single photons or packets of photons in the laser beam with the same group velocity can further be measured as a comb-like frequency spectrum, an experimental observation that was e. g. demonstrated in reference \cite{ref-4}. Since the group velocity of the wave packets is in principle not exactly equal to the phase velocity of the waves inside the wave packet, the famous comb frequency spectrum of coherent laser light is characterized by an offset frequency $\omega_0$, i. e. the measured light comb is shifted exactly by this offset in the frequency domain \cite{ref-5}. Frequency combs over a range of up to $20$ THz have been realized with an experimental uniformity of the repetition frequency of up to $10^{-19}$ \cite{ref-6}. Such high-precision measurements can be used to determine atomic level spectra with very high accuracy \cite{ref-7, ref-8}.           

In the development cycles and also after the achievement of the first Bose-Einstein condensates in the laboratory \cite{ref-10}, this new form of matter with the characteristics of long-range coherence in time and space has soon been associated with the scenario of coherent laser sources \cite{ref-9}. Follow-up experiments on interfering Bose-Einstein condensates have proven the ability of (atom-wise) constructive and destructive interference of ultra-cold coherent atomic vapors \cite{ref-11, ref-12, ref-13}, similar to the comb spectrum of coherent laser beams . In this context, the outcoupling of coherent atoms and ensembles of atoms from magneto-optical confinements is labeled atom laser in the scientific literature. An atom laser can be understood as a source of coherent atomic wave packets with well-defined phase relations between the different atomic matter waves below the critical temperature. In this respect, it is intuitive to calculate and model comb-like frequency spectra within a theoretical quantum field theory, similar to the Fourier concept of the coherent laser. However, in contrast to the case of a coherent laser beam which consists of a very large number of up to $10^{20}$ coherent photons, the phase coherence in the atomic vapor which normally consists of up to $10^3 - 10^5$ atoms is much more sensitive against thermal noise. For this reason, the standard Fourier ansatz of a coherent standing wave field neglecting noise induced by the thermal environment and finite size effects from particle-number conservation is not a sufficiently complex theoretical model to describe and explain the derivation of the frequency spectrum and the formal similarity of the latter to the comb spectrum of the coherent optical laser.     
  
To this end, a theoretical quantum field theory based on a Monte-Carlo sampling algorithm is presented in the framework of the present article, which illustrates the analytical derivation of the frequency comb spectrum for an atomic laser subject to thermal noise induced by the temperature of the surrounding atomic vapor, through the selection of quantum field components with multiples of the same phase difference entailed in the mathematically quantized field of the atom laser. In the sequel of this theory, the complex-valued quantum field and the resulting frequency comb distribution with the varying shape of its envelope is numerically modeled and monitored within a quantitative Monte-Carlo sampling method, as a function of temperature and external trap frequency. As an outstanding feature of this new formal approach as a part of the theory for atom lasers, it is explained how the finite particle number of the atomic vapor leads to the quantization of time evolution in units of the coherence time $\tau$ defined by the repetition frequency $\omega$ of the external driving laser field. Since the repetition frequency is not independent of the trap geometry of the external confinement for finite atomic samples, the partial chemical potentials of the atomic laser field define a distribution of possible values, as well as a natural lower and upper bound for the repetition frequency of the comb spectrum, i. e. the accuracy of the atom laser. The lower and upper bounds effectively vanish in the semi-classical limit of optical lasers with an ideally infinite number of atoms. In particular, it is observed that the repetition frequency for a constant spacing of phase repetition is a linear function of temperature, approximately independent of the total particle number.      

\section{Theory}

To derive the frequency spectrum of an atom laser realized by releasing coherent wave packets from a quantum field confined in an external three-dimensional external potential, one may define the time-dependent and spatially integrated quantum field in the complex plane as described in reference \cite{ref-1}. Such coherent quantum fields have e. g. been experimentally realized with a coherent laser beam in a microwave cavity in references \cite{ref-2, ref-3}, or based on atomic lasers in reference \cite{ref-14}. In the absence of external noise and fluctuations induced by temperature, the complex-valued quantum field can be divided into two different coherent field components with unique physical meaning, $\psi_+(t)$ a field component that propagates in the forward direction of time, and $\psi_-(t)$ the time-reversed counterpart. Fourier transformation of the intensity (absolute square of either forward propagating laser field $\psi_+(t)$, backward propagating field $\psi_-(t)$ or composite standing wave field $\Psi(t) = \frac{1}{2}(\psi_+(t) + \psi_-(t))$ of parallel forward and backward propagating optical laser fields with standard Fourier transformation \cite{ref-11} typically leads to a Gaussian-shaped frequency comb, which depends on phase shifts, frequency components and the intensity distribution of the quantum field in second quantization. In the standard approaches using the concept of mathematical Fourier transformation, typically, no further assumptions on the temperature dependence are applied, in the sense that the laser field is simply described by a pure quantum state in a complex plane (plane wave or wave field in an external confinement), rather than a statistical density matrix and no assumptions are made to describe noise effects induced by incoherent thermal mixing.  

In the present model, the quantum field of the atom laser confined in a microwave cavity characterized by trap frequencies $(\omega_x, \omega_y, \omega_z)$ and single particle energies $\epsilon_{\bf{k}} = \hbar(k_x\omega_x + k_y\omega_y + k_z\omega_z)$ is first of all defined in terms of a two-dimensional quantum vector which entails the orthogonal and independent forward and backward propagating field components $\psi_+(t)$ and $\psi_-(t)$ in the external confinement with trap frequencies $\omega_x$, $\omega_y$ and $\omega_z$, as long as no noise effects lead to the mixing (decoherence) of the orthogonal field components $\psi_+(t)$ and $\psi_-(t)$. In this case, in the absence of temperature, one may generally define the quantum field of an atom laser mathematically, as 

\begin{equation}
\label{eq.1}
\vec{\Psi}(t) =  \left(\begin{array}{c} \psi_+(t) \\ \psi_-(t) \end{array}\right) \ ,
\end{equation}\\
with time-dependent forward and backward propagating laser fields

\begin{equation}
\label{eq.2}
\psi_{\pm}(t) = \sum_{\bf{k}}c_{\bf{k}}{\rm e}^{\mp i\phi_{\bf{k}}(t)} . 
\end{equation}\\
Similar to the concept of Fourier series expansion, we further assert that the two field components $\psi_+(t)$ and $\psi_-(t)$ entail a symmetry property concerning a period $\tau$. This symmetry is thus represented by the time scale $t=\tau$ (coherence time) and may e. g. be realized by performing the outcoupling of partial atomic waves from the ultra-cold trapped vapor due to gravitational or electro-static forces at the edges of a magneto-optical cavity at multiples of the symmetry time interval, i. e. $t=m\tau$, with integer values of $m$, where 

\begin{equation}
\label{eq.5}
\psi_+(t=m\tau) = \psi_-(t=m\tau)\ .
\end{equation}\\
As we will further see, this symmetric ansatz leads to the same frequency comb spectrum as the standard Fourier expansion, which disassembles the total wave field of the atom laser into field components with integer multiples of the repetition frequency for time-periodic quantum fields. 

The mathematical phase in Eq. (\ref{eq.1}) can be directly related to the partial chemical potentials $\mu_{\bf{k}}$ of the atomic quantum field modes, through the relation $\phi_{\bf{k}}(t) = {\rm Re}\lbrace\frac{\mu_{\bf{k}}t}{\hbar}\rbrace$. The partial chemical potentials can be quantified by the conservation of the average total number of particles in the atomic laser field, from the fact that

\begin{equation}
\label{eq.3}
\sum^{\infty}_{j\ne0} z^{j}(\mu) \left[\prod_{l=x,y,z} \frac{1}{1-{\rm e}^{-j\beta\hbar\omega_l}} - 1\right] = (N-N_0) ,
\end{equation}\\
with $z(\mu)= {\rm e}^{\beta\hbar\mu}$ the fugacity of the atom laser and $\omega_{l}$ the photon frequency of the laser (in mode direction $l = x, y, z$). Because of the periodicity of the atomic laser field, this is the case, if

\begin{equation}
\label{eq.4}
\frac{\phi_{\bf{k}}(\tau)}{\tau} = \omega m({\bf{k}}) + \omega_0 ,
\end{equation}\\
where $\omega = \frac{2\pi}{\tau}$ is the (intrinsically derived) repetition frequency of the (real-valued) comb-like spectrum $\Omega = \lbrace\phi_{\bf{k}}\rbrace$ defined by the real parts of the quantized chemical potential $\mu_{\bf{k}}$, the coherence time $\tau$ and $\omega_0$ is the offset frequency. Note that the repetition frequency $\omega$ is in general not a multiple of the offset frequency $\omega_0$ and tends to zero in the semiclassical limit of vanishing chemical potentials. Fourier series expansion of the quantum field in Eq. (\ref{eq.2}) would lead to the same frequency comb spectrum in Eq. (\ref{eq.3}) through disassembling the total wave field into field components with integer multiples of the repetition frequency for time-periodic quantum fields, however, without physical and quantitative relation to the (partial) temperature-dependent chemical potentials $\mu_{\bf{k}}$. Since Eq. (\ref{eq.2}) entails a natural time unit, i. e. symmetry and quantization of time concerning the period $t=\tau$ (coherence time) after which the forward and backward propagating complex-valued quantum fields equal, it holds that the phases $m\phi_{\bf{k}} = m\omega_0\tau + 2\pi n$, which leads to the equality $\psi_+(t=m\tau) = \psi_-(t=m\tau)$ and finally to Eq. (\ref{eq.4}). 

Effects of thermal noise can be mathematically accounted for by relating the coherent time evolution of the quantum field with decoherence effects induced by temperature with the formal concept of imaginary time.
More precisely, in the case of coupling the coherent laser field to the noisy, incoherent part of the quantum field, the Wick rotation

\begin{equation}
\label{eq.6}
\tau = it = \beta\hbar
\end{equation}\\
formally accounts for the expression of the direction of time evolution, where $\beta$ denotes the inverse temperature and $\hbar$ the Planck constant. Since the coherent laser field is no longer isolated from its thermal environment, we observe that both field components $\psi_+$ and $\psi_-$ become effectively mixed and equal, i. e. in particular independent of the direction of time propagation (no time-reversal symmetry) at finite temperature, so that 

\begin{equation}
\label{eq.7}
\psi = \psi_{\pm} = \frac{1}{2}(\psi_{+} + \psi_{-}) = \sum_{\bf{k}}c_{\bf{k}}{\rm e}^{-\beta\mu_{\bf{k}}} . 
\end{equation}\\
The latter fact is accounted for by the (formal) equality $ it = \beta\hbar = -it$ which holds for the case of quantum fields with time-periodic symmetry concerning the coherence time $\tau$, where the field frequency is an integer multiple of the repetition frequency $\omega = \frac{2\pi}{\tau}$. The repetition frequency depends on the frequency of the external confinement, the number of particles occupying the different field modes, as well as on the temperature associated with the (thermalized) quantum field. For this reason, the partial chemical potential $\mu_{\bf{k}}$ is no longer purely real-valued, but a complex number that also entails the decay rate of the corresponding quantum mode in the coherent expansion of the atomic laser field. Please note that, in the case of non-periodic wave fields, simply replacing $it$ with $\beta\hbar$ would lead to a divergence in the time direction (forward and backward propagation) in the limit $T\rightarrow0^{+}$. The symmetry condition leads to a well-defined and unique quantization of time for the quantum field at finite temperatures (in units of the coherence time $\tau$ of the atom laser). Since the quantum field is assumed to be thermalized and follow a unique time direction, the result in Eq. (\ref{eq.7}) is independent of whether the field is initially assumed to be a standing wave or an orthogonal composition of two orthogonal field components $\psi_+$ and $\psi_-$, respectively.  

\begin{figure}
\begin{center}
\includegraphics[width=7.0cm, height=5.5cm,angle=0.0]{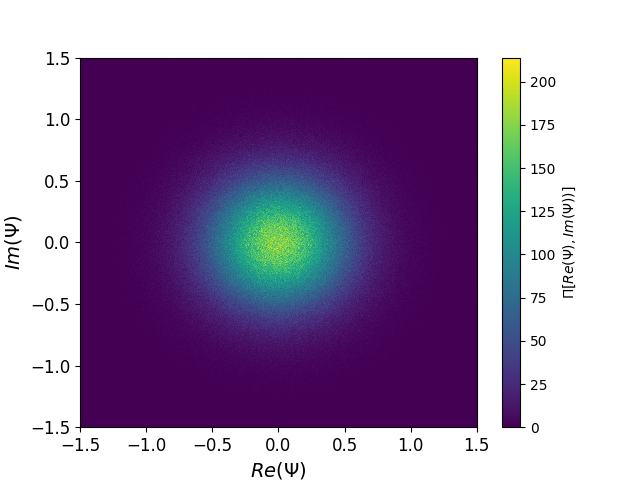} 
\caption{(Color online) Figure shows $5\times10^6$ realizations of the quantum field for an atom laser simulated from partial waves of a Bose-Einstein condensate in a harmonic trap at temperature $T=10{\rm~nK}$. According to Eq. (\ref{eq.7}) for the atomic sample, the atom laser field is represented in the complex plane. The external circular trap frequency is set to $\omega_x = \omega_y = \omega_z = 2\pi\times125 {\rm~Hz}$ and the average particle number to $\overline{N}=5\times10^{3}$.}
\label{fig.1}
\end{center}
\end{figure}

\section{Numerical Monte-Carlo modeling}
Given the numerical input parameters such as particle number $N$, temperature $T$ and external trap frequencies $(\omega_x, \omega_y, \omega_z)$, the analytic theory in section II for the derivation of the frequency comb for an atom laser can be applied for stochastic modeling of the relevant spectrum within a Monte-Carlo approach, as a function of the above model input parameters. Typical particle numbers of Bose-Einstein condensates range from a few thousand to millions of atoms, about fifteen orders of magnitude lower than the number of photons in an optical laser. At ultra-cold temperatures as low as a few hundred Nanokelvin, the atoms confined in the external potential of a magneto-optical trap tend to undergo the well-known phase transition into a Bose-Einstein condensate. The trap frequencies of the external confinement have been chosen in a few hundred Hz range for the present numerical Monte-Carlo simulations.

\begin{figure}[t]
\begin{center}
\includegraphics[width=4.25cm, height=4.25cm,angle=0.0]{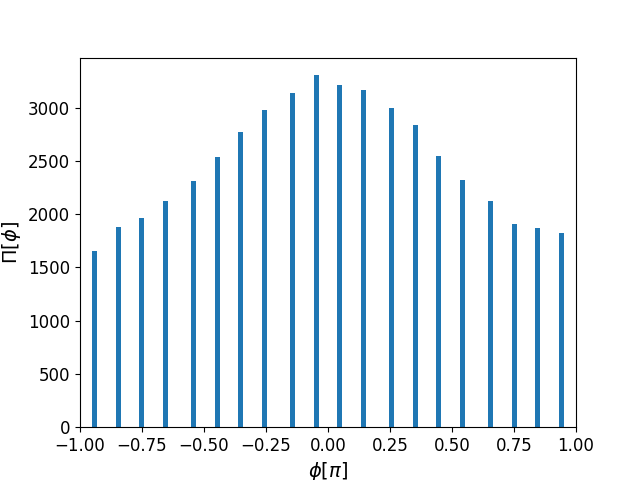} 
\includegraphics[width=4.25cm, height=4.25cm,angle=0.0]{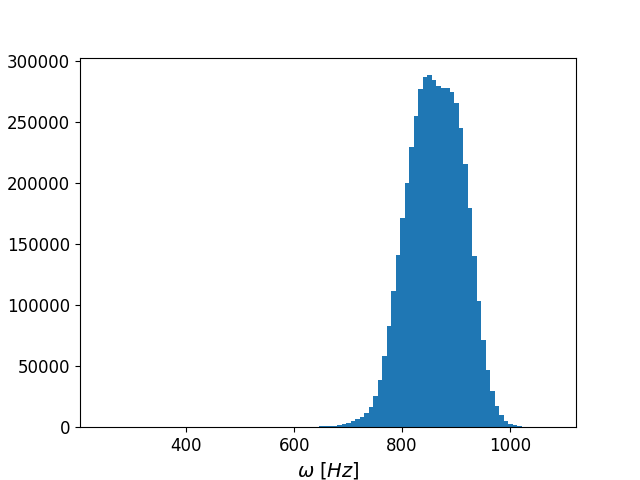} 
\includegraphics[width=4.25cm, height=4.25cm,angle=0.0]{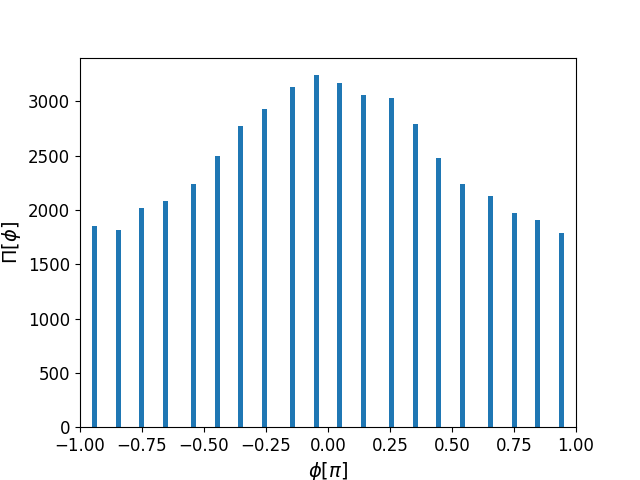} 
\includegraphics[width=4.25cm, height=4.25cm,angle=0.0]{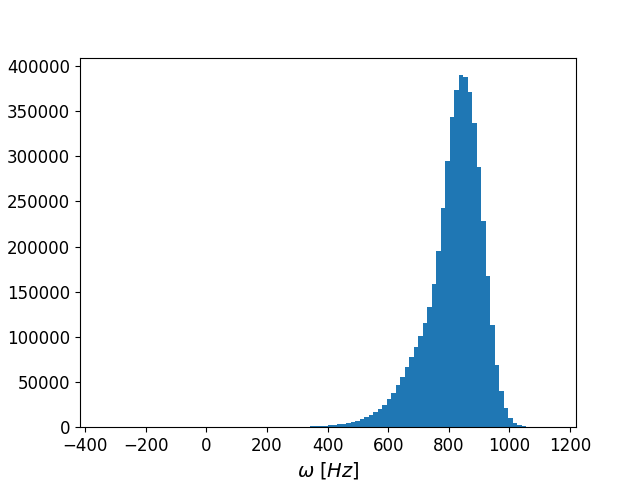}
\includegraphics[width=4.25cm, height=4.25cm,angle=0.0]{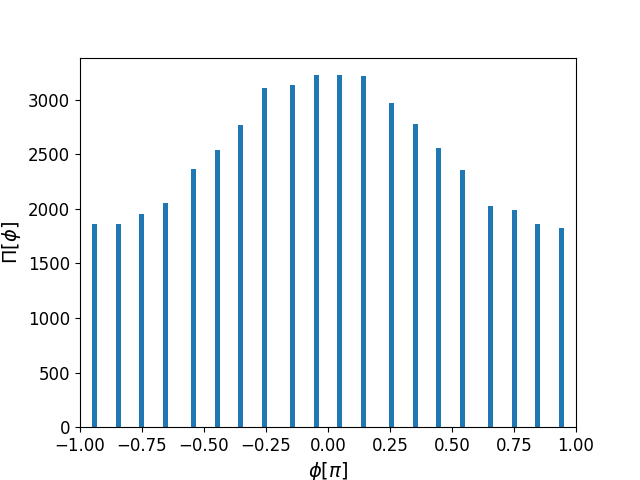} 
\includegraphics[width=4.25cm, height=4.25cm,angle=0.0]{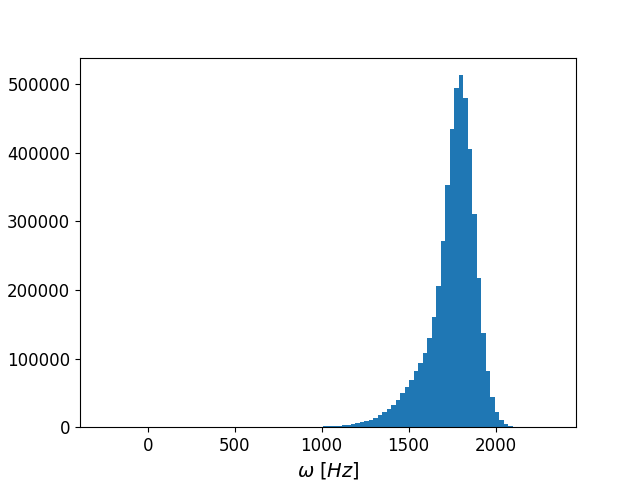} 
\caption{(Color online) Figure shows three different frequency phase comb spectra (repetition frequency $\omega$ times coherence time $\tau$) for a chosen unit phase of $\phi_0=\frac{\pi}{20}$ with corresponding repetition frequency distributions, as expected from Eq. (\ref{eq.4}). The spectra are obtained from numerical simulations with isotropic trap frequencies $\omega_x = \omega_y = \omega_z = 2\pi\times125 {\rm~Hz}$ (upper figure) and for non-isotropic frequencies $\omega_x = 2\pi\times125 {\rm~Hz}, \omega_y = 2\pi\times75 {\rm~Hz}, \omega_z = 2\pi\times25 {\rm~Hz}$ (middle figure) for the average particle number of $\overline{N}=5\times10^{3}$ and the same temperature of $T=25.0{\rm ~nK}$. As indicated by the results, the (average) repetition frequency $\omega$ mainly depends on the temperature of the atomic laser source (Bose-Einstein condensate) below the critical temperature of $T_c = 47.6{\rm ~nK}$. This is confirmed by the numerical simulation for a different temperature $T=50.0 {\rm~nK}$ (lower figure). The symmetry of the distribution for the repetition frequency changes to an asymmetrically shaped distribution for non-isotropic trap geometries.}
\label{fig.2}
\end{center}
\end{figure}   

To build the numerical calculation routine in the programming language Python, the quantum field in Eq. (\ref{eq.7}) is modeled for different realizations of the model parameters. The total particle number is assumed to be Gauss distributed around $\overline{N}$, temperature and the external trap frequencies have been kept constant in each cycling step of a simulation run. The partial chemical potentials $\mu_{\bf{k}}$ in Eq. (\ref{eq.7}) are obtained from the numerical solution of the intrinsic Eq. (\ref{eq.3}) for a given average total atom number $\overline{N}$ at temperature $T$ in the Bose gas. Finding poles of the Eq. (\ref{eq.3}) for the chemical potential for a set of model input parameters $(N, T, \omega_{\bf{k}})$ leads to a set of $M$ complex-valued partial chemical potentials $\Omega(\mu) = \Omega(\omega) + i\Omega(\Gamma) = \lbrace\mu_{\bf{k}}\rbrace$, where $M$ is the number of modes used to calculate the quantum field of the atom laser. Typical ranges of the effectively occupied number of eigenmodes in a Bose-Einstein condensate at ultra-cold temperatures consist of a few hundred modes, much less than the number of a few hundred thousand contributing modes in the case of an optical laser. Because of the symmetry of the quantum field for the atom laser, the sum of the imaginary parts of two respective component pairs of the atom laser field, shown for a given parameter set in Fig. (\ref{fig.1}), as well as the ensemble averages of the atom laser field $\psi$ vanishes.

\begin{figure}[t]
\begin{center}
\includegraphics[width=4.25cm, height=3.25cm,angle=0.0]{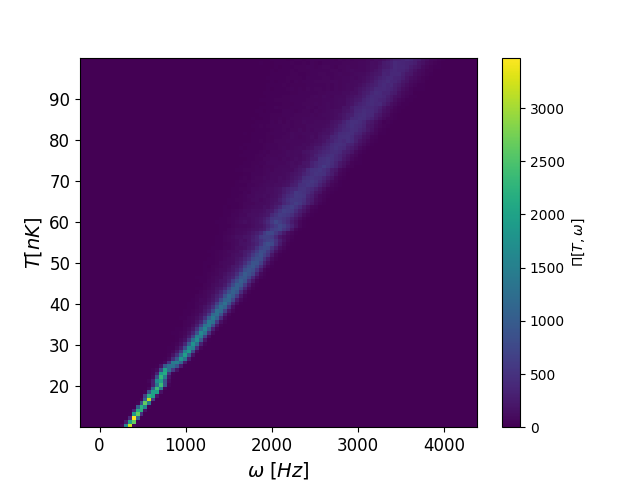} 
\includegraphics[width=4.25cm, height=3.25cm,angle=0.0]{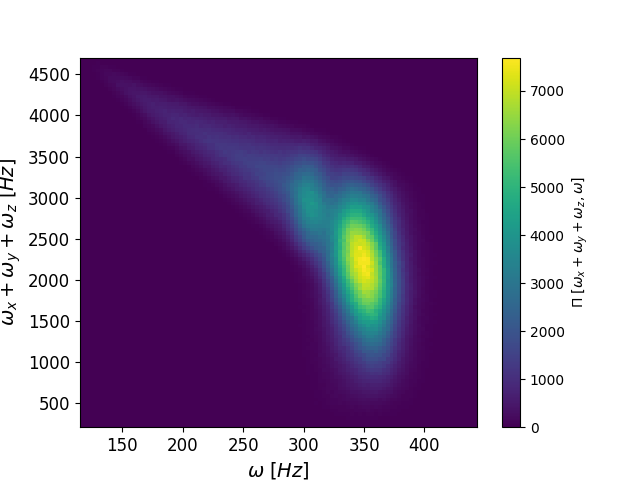} 
\includegraphics[width=4.25cm, height=3.25cm,angle=0.0]{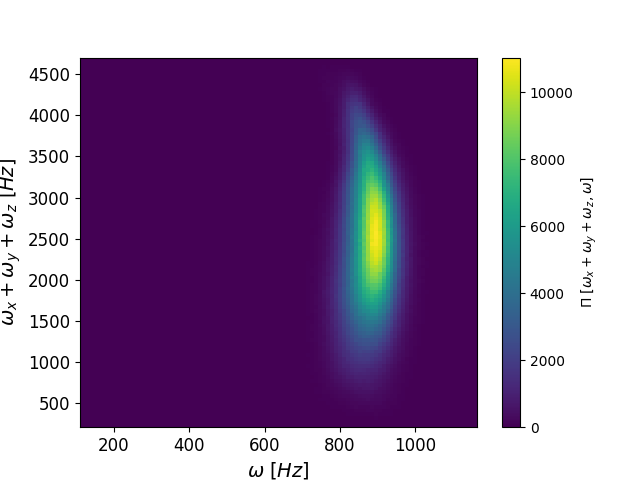} 
\includegraphics[width=4.25cm, height=3.25cm,angle=0.0]{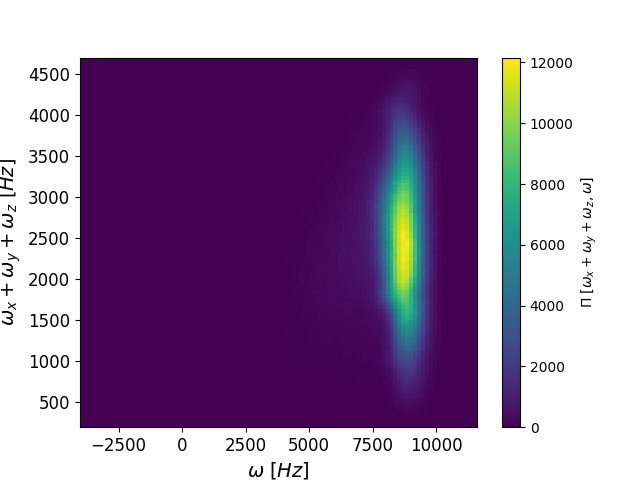} 
\caption{(Color online) The upper left figure shows the numerically calculated dependence of temperature and the repetition frequency from Eq. (\ref{eq.8}) with an approximately linear scaling behavior. The linear scaling is nearly independent of the external potential in the realistic parameter range of $(\omega_x, \omega_y, \omega_z)$, i. e. in the range of $2\pi\times10{\rm~Hz}$ to $2\pi\times250{\rm~Hz}$. The upper right, lower left and lower right figures further highlight the distribution of the repetition frequency as a function of the external trap frequency for different temperatures, $T=10{\rm~nK}$ (upper right), $T = 25 {\rm~nK}$ (lower left) and $T = 100{\rm~nK}$ (lower right). As the Monte-Carlo simulations indicate, the frequencies of the external confinement effectively define a maximal range for the distribution of the repetition frequency from the conservation of energy with the largest range for low temperatures. The density of the distribution patterns is increased for non-isotropic trap geometries concerning isotropic external confinements by approximately a factor of five (not shown).}
\label{fig.3}
\end{center}
\end{figure}  

To calculate the frequency comb spectrum $\Omega(\omega) = {\rm Re}\lbrace\Omega(\mu)\rbrace = \lbrace\omega m(\bf{k}) + \omega_0\rbrace$ for realizations of the quantum field in Eq. (\ref{eq.7}) with different model parameters of the temperature $T$ and particle number $N$ at given constant frequencies $(\omega_x, \omega_y, \omega_z)$ of the external confinement using a Monte-Carlo sampling technique, the total phase of the quantum field for the atom laser is calculated numerically, given that

\begin{equation}
\label{eq.8}
\psi = \vert\psi\vert{\rm e}^{i\phi} = \sum_{\bf{k}}c_{\bf{k}}{\rm e}^{-\beta\mu_{\bf{k}}} \ .
\end{equation}\\
Hence, the solution of Eq. (\ref{eq.8}) for the phase $\phi=\beta\mu=\beta\hbar\omega m(\bf{k})+\beta\omega_0$ is obtained from the partial chemical potentials $\mu_{\bf{k}}$ and the mode weighting factors $c_{\bf{k}}$, which define the various coherent phase relations between the eigenmodes of the condensed particles in the external harmonic trap in the presence of the thermal atomic vapor. Since the Monte-Carlo sampling algorithm is iterated over the ratio of exponential probabilities described by the Boltzmann factor ${\rm exp}(-\beta\vert\mu\vert)$, transition probabilities from quantum field states with different quantization number $m(\bf{k})$ exponentially decay, so that from the initial condition that $\beta\mu=1.0$ and numerical calculation of the distribution of $\mu$ from Eq. (\ref{eq.8}) one effectively obtains the distribution of the repetition frequency $\omega$ (with $\omega_0 = 0.0$ due to the symmetry of the atom laser comb), as shown in figures (\ref{fig.2}) and (\ref{fig.3}). Measurements (or outcoupling) of quantized phase shifts $\Delta\phi = k\omega\tau + \omega_0\tau$ are numerically simulated by collecting only realizations with integer multiples of a certain (self-defined) phase shift $\phi_0$ in the Monte-Carlo simulation of the atom laser field. For the simulations represented in figures (\ref{fig.1}) and (\ref{fig.2}), the phase resolution $\phi_0$ was set to $\frac{\pi}{20}$. 

For the atom laser described in the present theory, the highest phase resolution $\phi_0$ is limited by the quantization of the chemical potential (and time equivalently), as defined by Eq. (\ref{eq.4}), since the total particle number is finite. In comparison, the precision of an optical laser can not be approached in reality, but only formally in the semi-classical limit of infinite particle numbers, since the number of particles in the largest Bose-Einstein condensates so far realized in the experiment is on the order of $10^{6}-10^{7}$. The distribution of the repetition frequency is numerically calculated by counting realizations of the repetition frequency as a function of temperature for different values of the temperature, and a linear scaling behavior is observed for the typical range of the total average number of particles in the harmonic optical trap. In dependence on the external trap frequency, the distribution of the repetition frequency is also calculated for different realizations of the external trap frequencies $(\omega_x, \omega_y, \omega_z)$ in the range from $2\pi\times10{\rm~Hz}$ to $2\pi\times250{\rm~Hz}$ for three different temperatures $T=10{\rm~nK}, 25{\rm~nK}$ and $100{\rm~nK}$.   

\section{Discussion}

In our present theory, the quantum field of an atomic laser at finite temperatures in Eq. (8) builds the starting point for the numerical analysis of an atomic laser. From Eq. (8), the quantum field is modeled calculating the realizations of the field with random numbers $c_{\bf{k}}$ that do represent the weights of the field component with different complex frequencies as defined by the partial chemical potentials $\mu_{\bf{k}}$ for the different field modes. At finite temperatures below the critical temperature, the partial (mode dependent) chemical potentials $\frac{\mu_{\bf{k}}}{\hbar} = \omega_{\bf{k}} + i\Gamma_{\bf{k}}$ do exhibit finite decay rates $\Gamma_{\bf{k}}$ besides the real-valued coherent (real) parts $\omega_{\bf{k}}$ of the mathematical phases $\phi_{\bf{k}}$. In the complex plane, which is a suitable mathematical representation to understand the composition and structure of the quantum field for an atom laser, finite decay rates (obtained from the imaginary parts of the chemical potentials $\mu_{\bf{k}}$) lead to the shift of the quantum field distribution towards the zero point axis both for the condensed and the non-condensed part of the atomic laser built from the Bose-Einstein condensate, which is represented by different parts of the quantum field. In the present approach, the coherent sum of both fields (condensate and non-condensate part) of the quantum field is represented and builds the mathematical foundation of the atom laser. Since the average of the total quantum field is zero and the coherent phases lead to $\psi_+(t=m\tau) = \psi_-(t=m\tau)$ on the time scale for coherent particle interactions $\tau$ (field symmetry in the complex plane), the total field of the atom laser as represented in Fig. (\ref{fig.1}) is independent of the decay characteristics described by $\Omega(\Gamma)={\rm Im}\lbrace\Omega(\mu)\rbrace = \lbrace\Gamma_{\bf{k}}\rbrace$.    

The real part of the frequency spectrum entails frequency components in quantized units of a constant phase or repetition frequency, respectively. Since the chemical potential and therefore the repetition frequency (and time) is quantized, the numerically derived frequency comb spectrum from Eq. (\ref{eq.4}) of the present quantum field theory is thus limited in accuracy. In comparison to the optical laser, we observe that the minimal repetition frequency in units of the time scale $\tau$ is about $10\%$ of the external trap frequency. Simulation of the quantum field for realizations with a phase difference of a constant $\phi_0$ is shown in Fig. (\ref{fig.2}) for the case $\phi_0=\frac{\pi}{20}$. From this Monte-Carlo simulation, a shaped comb-like frequency spectrum is observed. As numerically verified, simulations of the frequency comb for different frequencies of the external potential show that the (Gaussian) shape of the frequency comb is approximately independent of the external confinement of the atomic cloud. As concerns the distribution of the repetition frequency, it is observed that the indicated symmetric Gaussian distribution for isotropic trap geometries slightly changes its symmetry property with an increased slope towards higher values of the repetition frequency and a decreased slope towards lower values of the repetition frequency. As further indicated in the upper left panel of Fig. (\ref{fig.3}), the repetition frequency is a linear function of temperature $T$.  

Although it can be shown and numerically verified that the atom laser and the standard optical laser obey the same type of comb-like frequency spectrum within the present Monte-Carlo simulation method for the frequency comb spectrum of an atom laser, it appears that the optical laser and the atom laser built from a Bose-Einstein condensate by releasing coherent atomic matter waves from the external trap do exhibit a fundamental difference. Whereas the atom laser mimics an optical laser by the outcoupling of single atoms or coherently formed wave packets of atoms from the trapped Bose-Einstein condensate which induces changes in energy as represented by the chemical potential, the photons of an atom laser do not induce a chemical potential and consequently effectively do not define a limitation to the measurement range for calibration measurements (with a possible range of the repetition frequency) of the frequency comb. For an optical laser, the accuracy of calibration measurements is limited basically by the natural line width of the laser frequency, whereas for an atom laser, it is the non-vanishing, quantized chemical potential on the order of the external trap frequencies that defines the accuracy of measurements performed with the atom laser. In the limit of large temperatures and infinite particle numbers, this limitation of the atom laser tends to vanish for very weakly or almost non-interacting, coherently coupled atoms which then follow the physical mechanisms and repetition frequency range of a standard optical laser.       

\section{Conclusion}

In conclusion, a new method based on a Monte-Carlo simulation approach for calculating frequency combs for atomic lasers built from releasing single atoms out of a trapped Bose-Einstein condensate at ultra-cold temperatures has been presented in the sequel of this article. It was shown how the comb-like frequency spectrum can be derived analytically assuming the quantum field of a Bose-Einstein condensate to obey a symmetry property concerning the field oscillation period and coherence time $\tau$. Using the concept of complex time, it is possible to switch from the time domain to representations of the combined condensate and non-condensate quantum field of the atom laser at finite temperature, accounting for the thermal mixing of different atom laser modes. Simulating a measurement of the symmetric combined condensate and non-condensate quantum field at finite temperature by collecting the relevant parameters of the numerical simulation only at definite mathematical (coherent) phases of the field leads to a comb-like frequency spectrum. To derive the corresponding frequency spectrum and its distribution, as well as the scaling of the repetition frequency with temperature, the scaled phase distribution is calculated by simulating the quantum field of the atom laser and deriving statistical distributions for the repetition frequency from the numerical simulation. 
        
\acknowledgments
The author acknowledges the financial support from IU Internationale Hochschule for the continuous and ongoing (freelancer) lecturer position at the university, which has in particular enabled the formulation and editing of the present theory 
on a Monte-Carlo simulation method for calculating the frequency comb of an atomic laser obtained from a Bose-Einstein condensate.

\end{document}